# Control of biaxial strain in single-layer Molybdenite using local thermal expansion of the substrate

*Gerd Plechinger[1], Andres Castellanos-Gomez[2], Michele Buscema[2], Herre S. J. van der Zant[2],*

*Gary A. Steele[2], Agnieszka Kuc[3], Thomas Heine[3], Christian Schüller[1], and Tobias Korn[1]\**

[1] Institut für Experimentelle und Angewandte Physik, Universität Regensburg, D-93040 Regensburg, Germany

[2] Kavli Institute of Nanoscience, Delft University of Technology, Lorentzweg 1, 2628 CJ Delft, The Netherlands.

[3] School of Engineering and Science, Jacobs University Bremen, Campus Ring 1, D-28759 Bremen, Germany.

tobias.korn@physik.uni-regensburg.de

ABSTRACT

Single-layer $MoS_2$ is a direct-gap semiconductor whose electronic band structure strongly depends on the strain applied to its crystal lattice. While uniaxial strain can be easily applied in a controlled way, e.g., by bending of a flexible substrate with the atomically thin $MoS_2$ layer on top, experimental realization of biaxial strain is more challenging. Here, we exploit the large mismatch between the thermal expansion coefficients of $MoS_2$ and a silicone-based substrate to apply a controllable biaxial tensile strain by heating the substrate with a focused laser. The effect of this biaxial strain is directly observable in optical spectroscopy as a redshift of the $MoS_2$ photoluminescence. We also demonstrate the potential of this method to engineer more complex strain patterns by employing highly absorptive features on the substrate to achieve non-uniform heat profiles. By comparison of the observed redshift to strain-dependent band structure calculations, we estimate the biaxial strain applied by the silicone-based substrate to be up to 0.2 percent, corresponding to a band gap modulation of 105 meV per percentage of biaxial tensile strain.

KEYWORDS

single-layer molybdenum disulfide, atomically thin crystal, strain engineering, bandstructure, biaxial strain

**Introduction**

Two-dimensional crystal structures which can be mechanically exfoliated from bulk layered crystals have attracted a lot of research interest in recent years. Beyond graphene, the stimulus of the field, a large variety of materials is available,[1] including semiconductors, superconductors[2] and topological insulators.[3] Among the semiconductors, dichalcogenides, such as $MoS_2$ and $WS_2$, are of special interest, as they show a transition from an indirect band gap in the bulk to a direct gap as a single layer.[4-6] These two-dimensional crystals have demonstrated outstanding mechanical properties, withstanding very large deformations before rupture, triggering interest in using strain engineering to modify their optoelectronic properties. The effect of uniaxial tensile strain on the band structure of $MoS_2$ has been extensively studied by density functional theory(DFT) [7-10] and by photoluminescence (PL) experiments. [11-14] The effect of biaxial strain on the band structure, on the other hand, is expected from recent calculations to be significantly stronger than that of uniaxial straining. [8,9,15,16] Local biaxial straining has also been recently proposed to generate a funnel for excitons, with interest in photovoltaics and photodetection applications. [17] However, experimental application of large biaxial strain is much more challenging than uniaxial straining and typically utilizes electromechanical devices based on piezoelectrics.[18] As a result, the amount of experimental works on biaxial strain engineering in $MoS_2$ is still very scarce and so far limited to multilayer $MoS_2$.[19]

Here, we demonstrate a simple way to apply *biaxial tensile* strain to single-layer $MoS_2$ by profiting from the large mismatch between the thermal expansion coefficients of $MoS_2$ and a polydimethylsiloxane (PDMS) substrate. The temperature of the PDMS substrate is changed globally by heating the whole sample or locally by a focused laser beam. The effect of the applied strain is directly seen as an observed redshift of the A exciton peak position in PL measurements. From the redshift of the exciton peak, we infer that the band gap of the material is shifting by up to 60 meV in response to the local heating. To extract the strain that can be applied, we compare the experimental results to DFT calculations and find that the local heating can yield up to 0.2 percent biaxial strain through thermal expansion of the PDMS substrate. Additionally, we demonstrate that laser-induced heating of multilayer $MoS_2$ flakes on top of PDMS can be used to create uniaxial *compressive* strain at a distance of several µm from the heating laser focus.

**Results and discussion**

MoS$_2$ flakes are prepared on PDMS and SiO$_2$ (used as reference sample) by mechanical exfoliation[20] (see experimental section). As evident from Figure 1a, single-layer regions of the flakes on PDMS are weakly visible in brightfield optical microscope images. The transparent PDMS substrates allow the use of optical microscopy in transmission mode (see Figure 1b), and the absolute value of the optical absorption can directly be extracted from transmission images. As Figure 1c shows, a single layer of MoS$_2$ on PDMS absorbs about 2 percent of the incident light in the green spectral range. This is in a good agreement with recent findings of universal interband absorption quanta in two-dimensional semiconductors.[21] Note that the large optical contrast (up to 20%) found by quantitative optical microscopy on single-layer MoS$_2$ on top of SiO$_2$/Si substrates is due to interference effects and multiple reflection/absorption processes.[22-24]

The combination of optical microscopy with PL provides a reliable identification of single layers of MoS$_2$. [11,12] For instance, while a single-layer of MoS$_2$ is a direct gap semiconductor, its multilayered counterparts present an indirect band gap. This leads to a strongly enhanced PL emission of the single-layers which can be used to distinguish them from multilayer flakes. Figure 1d shows a typical PL spectrum measured for a single-layer MoS$_2$ at room temperature, showing a strong emission peak (A exciton) which corresponds to a direct transition at the *K* point of the Brillouin zone from the conduction band minimum to the uppermost valence band maximum. [4,5] The B exciton, corresponding to the transition to the lower valence band maximum, is only weakly visible in the PL spectra due to the fast intra-valence-band relaxation processes. [25]

In our experiments, SiO$_2$ is used as a reference substrate due to its very low thermal expansion coefficient and poor adhesion to the MoS$_2$. In fact, previous low-temperature PL measurements demonstrated that the thermal expansion of MoS$_2$ is not inhibited if it is deposited on top of SiO$_2$ due to the weak MoS$_2$/SiO$_2$ adhesion.[26]

For temperature-dependent measurements, the samples are mounted on the surface of a heater inside a small vacuum chamber under low-pressure He atmosphere. We observe a redshift of the A exciton PL peak position and a corresponding band gap reduction with increasing temperature, as shown in Figure 2a. A direct comparison of the shifts measured on PDMS substrates and reference $SiO_2$ substrates is presented in Figure 2b, where measurements of 3 different single-layer $MoS_2$ flakes on PDMS are shown. The dashed lines indicate linear fits to the data. For the reference sample on the $SiO_2$ substrate, we can attribute the band gap reduction solely to the thermal expansion of the $MoS_2$ flake as the $SiO_2$ thermal expansion is much smaller than that of $MoS_2$ and the $SiO_2/MoS_2$ adhesion is very weak. For $MoS_2$ on top of PDMS, on the other hand, we find a larger band gap reduction as a function of temperature, see Table 1 and Figure 2b. From this observation, we infer that, apart from the $MoS_2$ lattice expansion, heating induces an additional biaxial tensile strain on the $MoS_2$ flake due to the large thermal expansion coefficient of the PDMS substrate. This additional strain is reproducible in subsequent heating cycles, and its magnitude remains constant on the timescale of the experiment, so that we may exclude slippage of the strained $MoS_2$ on top of the PDMS, indicating strong adhesion between the two materials. Comparison of the temperature-induced band gap shifts on the different substrates allows us to determine that about 40 percent of the observed band gap reduction, while increasing the temperature of $MoS_2$ on PDMS is due to biaxial strain induced by the thermal expansion mismatch between the $MoS_2$ and substrate. Figure 2c shows a schematic of the different temperature-induced effects on the PDMS and $SiO_2$ substrates.

We now discuss the laser-induced generation of local biaxial strain. These measurements are carried out at room temperature under ambient conditions. Figure 3a is a visualization of the experiment. The focused laser is partially absorbed in the single $MoS_2$ layer, and some of the absorbed power is transferred to the substrate, leading to a localized thermal expansion of the substrate region beneath the $MoS_2$. Here, we can neglect direct heating of either the $SiO_2$ or the PDMS by the laser, given that their optical absorption is significantly weaker than that of the $MoS_2$. As Figure 3b shows, increasing the laser excitation density leads to a pronounced redshift of the A exciton PL peak position. We note that the excitation densities used in the experiment depicted in Figure 3b are significantly lower than the values used for laser-induced ablation of few-layer $MoS_2$ on $SiO_2$ substrates (several GW cm$^{-2}$),[27] and also lower than the threshold for destruction of single-layer $MoS_2$ on PDMS (about 600 kW cm$^{-2}$), which will be further discussed below.

In Figure 3b, data from four different flakes on PDMS substrates is shown, and the dashed lines correspond to linear fits to the data. We clearly see that the laser-induced redshift on the PDMS substrate is about 3 times larger than on the $SiO_2$ substrate (see Table 1), yielding redshifts of up to 30 meV in the excitation density range investigated in this measurement series. The dependence of the PL redshift on sample temperature (shown in Figure 2) allows us to calibrate the laser-induced local temperature increase as a function of excitation density, the results are depicted in Figure 3c. We find that on the PDMS substrate, the temperature increases by about 54 K for an excitation density of 200 kW cm$^{-2}$. The same excitation density only yields an increase of 28 K on the $SiO_2$ substrate. As discussed above, the optical absorption of $MoS_2$ on PDMS is about an order of magnitude smaller than on $SiO_2$. Given that $SiO_2$ and PDMS have very comparable heat capacity per volume (see Table S2 in the Supporting information), the larger temperature increase on the PDMS substrate must stem from its low thermal conductance, which is an order of magnitude lower than that of $SiO_2$, reducing the heat sinking to the surrounding material . The temperature increase and corresponding biaxial strain induced by the laser excitation appears to be homogeneous within the illuminated area, as evidenced by the PL linewidth, which is comparable to that for a sample temperature corresponding to a similar redshift. We are able to quantify the biaxial strain applied by the PDMS substrate to the $MoS_2$ flake by performing strain-dependent DFT calculations of the band structure, which will be discussed below. The results are shown in Figure 3d: in the accessible excitation density range, up to 0.2 percent of biaxial tensile strain can be reached.

To test the limits of laser-induced redshifts, we perform a second measurement series with larger excitation densities; the results are summarized in Figure 4. We observe that with increasing excitation density, the A exciton PL redshifts by up to 60 meV (see Figure 4a) before saturation occurs. In these measurements, the exposure time for each excitation density is kept constant at 20 seconds. Thus, the integrated PL intensities obtained in the measurements are directly comparable. We observe that the PL intensity increases linearly up to an excitation density of at least 600 kW cm$^{-2}$, indicating that no saturation or damage to the $MoS_2$ occurs (see Figure 4b). For larger excitation densities, the PL intensity saturates and finally decreases, indicating laser-induced damage of the $MoS_2$ layer. After the measurement series, optical microscopy confirmed that the investigated flake was destroyed at the position of the laser focus. We note that the critical excitation density for $MoS_2$ on top of PDMS is significantly smaller than that for $SiO_2$ substrates, even though the optical absorption is weaker. This clearly indicates that $SiO_2$ substrates provide more effective heat sinking.

In order to extract the effective applied strain from the observed redshift, we compare our results to the DFT calculations. The effect of strain on the band structure of $MoS_2$ has been investigated by several groups, and the values reported for the reduction of the direct band gap as a function of biaxial tensile strain are on the order of 100 meV per percent, while the indirect band gap that develops for larger biaxial tensile strain shows a larger reduction with strain.[8,9, 15-17,28-31] These calculations are based on DFT, even though band gaps are typically underestimated at this level of theory. However, in the case of two-dimensional materials, in particular for $MoS_2$, a fortuitous error cancellation yields to a nearly perfect match of DFT results with experiment: while the band gap is in fact about 1 eV larger, 2D structures have a substantial exciton binding energy that needs to be considered, which incidentally is on the order of ~1 eV, thus yielding values close to the DFT results. [32]

Recent DFT calculations have shown that there is no electronic coupling between $MoS_2$ and a $SiO_2$ substrate,[33] therefore we only consider the $MoS_2$ monolayer in our calculations. The electronic interaction of $MoS_2$ with the PDMS substrate is also very weak, as demonstrated by photoluminescence experiments.[20] Details of the band structure calculations are outlined in Supporting Information and accompanied by detailed band structure plots for $MoS_2$ in equilibrium position and under 2.2 percent compression and expansion, both biaxial as well as uniaxial along armchair and zigzag directions. Figure 5 shows the evolution of the band gap under biaxial and uniaxial strain in the range of 2 percent of compression and expansion. Within the investigated range, the band gap evolution is almost perfectly linear. At uniaxial compression of ~0.5 percent, discontinuities are observed both for compression along armchair and zigzag directions. The optical band gap change under biaxial strain is linear over the entire range investigated in the calculations and changes with 105 meV per percent of strain (see Table S1). However, under compression, $MoS_2$ quickly changes to become an indirect band gap semiconductor, and the fundamental gap never significantly exceeds the value of the equilibrium structure.

With the calculated band gap reduction of 105 meV per percent of strain, we can conclude that local optical excitation enables us to generate a local biaxial strain of at least 0.2 percent (corresponding to 40 percent of the observed band gap reduction of 60 meV) for a single-layer MoS$_2$ on PDMS. This value is lower than one might expect from the large difference of the thermal expansion coefficients, but we also have to consider the large mismatch in the Young's moduli of PDMS and MoS$_2$. Recent measurement on two-dimensional crystals demonstrated that the Young's modulus of MoS$_2$ is very high[34] (about 270-330 GPa), in the same order of magnitude as that of graphene[35,36] (about 1TPa). By contrast, the Young's modulus for PDMS is only about 0.8 GPa. Therefore, even a monolayer MoS$_2$ flake may limit the lateral thermal expansion of the PDMS substrate.

In the experiments discussed above, the strain imparted by the PDMS substrate to the MoS$_2$ is accompanied by a substantial temperature increase. In order to investigate if these two effects can be decoupled, we perform experiments using spatially separated laser focal spots for heating and strain probing. For this, a second laser is coupled into the microscope objective and focused onto the sample at a point 10 μm away from the PL excitation spot. This second laser spot is positioned on the few-layer region of a MoS$_2$ flake attached to the single-layer part probed by the PL excitation. We utilized a weak, constant excitation density of 26 kW cm$^{-2}$ for PL excitation, and varied the excitation density of the second (heating) laser.

Focusing the second laser onto a few-layer MoS$_2$ flake attached to the single-layer MoS$_2$ leads to a pronounced *blueshift* of the A exciton (about 7 meV for an excitation density of 1000 kW cm$^{-2}$), as shown in Figure 6a. This effect is absent if the second laser is focused onto the bare PDMS substrate (see Supporting Information). Our DFT calculations show that such a blueshift and corresponding *increase* of the band gap can only be caused by *compressive* strain acting on the MoS$_2$. We can explain this observation as follows: the second laser is strongly absorbed in the few-layer MoS$_2$, which in turn *locally* heats the PDMS substrate below. This local heating leads to thermal expansion of the PDMS, which is confined to a small volume due to the small thermal conductance of the substrate. The substrate surrounding that volume is compressed, resulting in near-uniaxial *compressive* strain acting on the single-layer part of the MoS$_2$ flake next to it, see Figure 6b for a schematic of this effect. This strain increases the band gap and leads to a blueshift of the A exciton. We can estimate the magnitude of the strain by comparing our observation to theory calculations.

As shown in Figure 5, uniaxial compression of up to 0.5 percent increases the band gap by 57 meV per percent. However, further compression will reverse the trend and lead to an anisotropic band gap reduction. Our values indicate that the sample was compressed by about 0.12 percent. This method of generating uniaxial compressive strain may be of interest for studies where local laser heating would strongly perturb the experiment due to the large density of optically generated photocarriers. In principle, complex local strain patterns can be generated using this method by controlled deposition or microstructuring of highly absorptive few-layer $MoS_2$ on PDMS substrates.

**Conclusions**

In summary, we exploited the large thermal expansion of PDMS substrates to apply controlled biaxial strain to a single-layer $MoS_2$. By studying the temperature-dependent shifts of the A exciton peak position on different substrates, we infer that the PDMS substrate exerts a temperature-dependent, biaxial tensile strain on the $MoS_2$. This strain can also be applied locally, using laser heating. For this, we vary the excitation density of the laser used to excite the photoluminescence. We observe a three times larger redshift of $MoS_2$ on top of PDMS as compared to $SiO_2$ substrates. This large shift is due to the small thermal conductivity of the PDMS, which leads to a larger local heating. Comparison of the observed redshift of the PL to the band structure calculations allows us to estimate the biaxial tensile strain that is applied to the $MoS_2$. We find a value of at least 0.2 percent. Additionally, by using a second laser to locally heat the PDMS substrate beneath highly absorptive multilayer $MoS_2$, we are able to create uniaxial compressive strain of about 0.1 percent on a single-layer $MoS_2$ flake at a distance of 10μm from the laser focus. These simple methods for local application and optical readout of uniaxial and biaxial strain may be applicable to a large number of two-dimensional crystal structures which have optical absorption in the visible range.

**Materials and methods**

*Sample preparation*

We prepared $MoS_2$ flakes on elastomeric substrates (Gel-Film® WF 6.0mil ×4 films) by mechanical exfoliation of natural $MoS_2$ (SPI Supplies, 429ML-AB) with blue Nitto tape (Nitto DenkoCo., SPV 224P). Gel-Film® is a commercially available elastomeric film based on polydimethylsiloxane (PDMS). $MoS_2$ flakes used for comparison are prepared by mechanical exfoliation on a p-doped silicon wafer with 285 nm $SiO_2$ thermal oxide on top.

*Photoluminescence*

For photoluminescence (PL) measurements, we utilize a microscope setup, in which a 532 nm cw laser is coupled into a 100x (1 μm laser spot size) microscope objective, which also collects the light emitted from the sample in backscattering geometry. The PL is recorded using a grating spectrometer equipped with a Peltier-cooled charge-coupled device (CCD) sensor. A longpass filter is placed at the entrance of the spectrometer to suppress elastically scattered laser light. The sample is mounted on a motorized XY table and can be scanned under the microscope to position the laser spot. Power-dependent PL measurements are carried out under ambient conditions. For temperature-dependent measurements, the samples are mounted on the surface of a heater inside a small vacuum chamber. The chamber is first evacuated and subsequently filled with He gas at a pressure of about 5 mBar.

*Spatially separated heating and strain probing*

The experiments on spatially separated heating and strain probing were carried out in ambient conditions. A second (heating) laser (532 nm, cw operation) was coupled into the microscope objective of the setup described above. It was focused (spot size about 1μm) onto the multilayer part of a $MoS_2$ flake, at a point 10 μm away from the PL excitation laser, which was positioned on the single-layer part of a $MoS_2$ flake. The laser used for PL excitation was kept at a constant, low excitation density of 26 $kWcm^{-2}$. In each measurement series, PL spectra were collected for various excitation densities of the heating laser.

*Band structure calculations*

All density-functional calculations have been carried out with the help of the ADF/BAND code[37] using explicit two-dimensional boundary conditions employing numerical and Slater-type basis functions of valence triple zeta quality with one polarization function (TZP)[38] with a small frozen core, the PBE-D3(BJ) functional. [39,40]

## ACKNOWLEDGMENT

The authors acknowledge financial support by the DFG via SFB689, KO3612/1-1,GRK 1570, HE3543/18-1 and the Dutch organization for Fundamental Research on Matter (FOM). A.C.-G. acknowledges financial support through the FP7-Marie Curie Projects PIEF-GA-

2011-300802 ('STRENGTHNANO'). A.K. and T.H. acknowledge financial support through FP7-MC-2012-ITN (GA 317451) ('MoWSeS').

**Figures and Tables**

| Substrate | Heat-induced band gap shift (meV/K) | Excitation-induced band gap shift (meVcm$^2$/kW) |
|---|---|---|
| SiO$_2$ | -0.295 | -0.041 |
| PDMS | -0.42 | -0.114 |

**Table 1.** Temperature- and excitation-density-induced band gap shifts of MoS on PDMS and SiO$_2$.

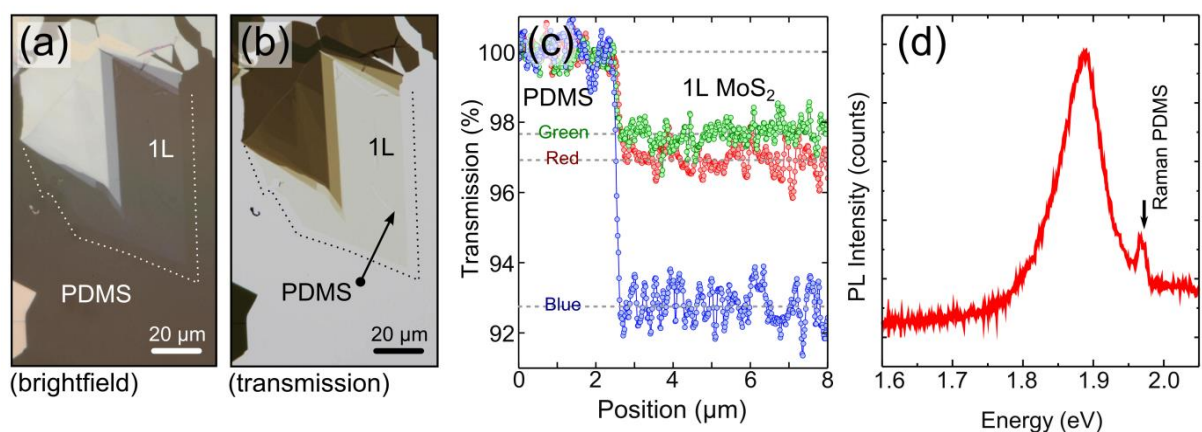

**Figure 1. Characterization of MoS$_2$ on PDMS substrate**. (a) Brightfield optical microscope image of MoS$_2$ flakes on PDMS. The dotted outline indicates a single-layer region. (b) Transmission optical microscope image. The black arrow indicates the position of the linescan shown in (c). (c) Normalized transmission traces extracted from the green, red and blue channels of (b). (d) Typical PL spectrum collected on single-layer MoS$_2$ on PDMS at room temperature using low excitation density. The black arrow indicates a Raman signal of the substrate.

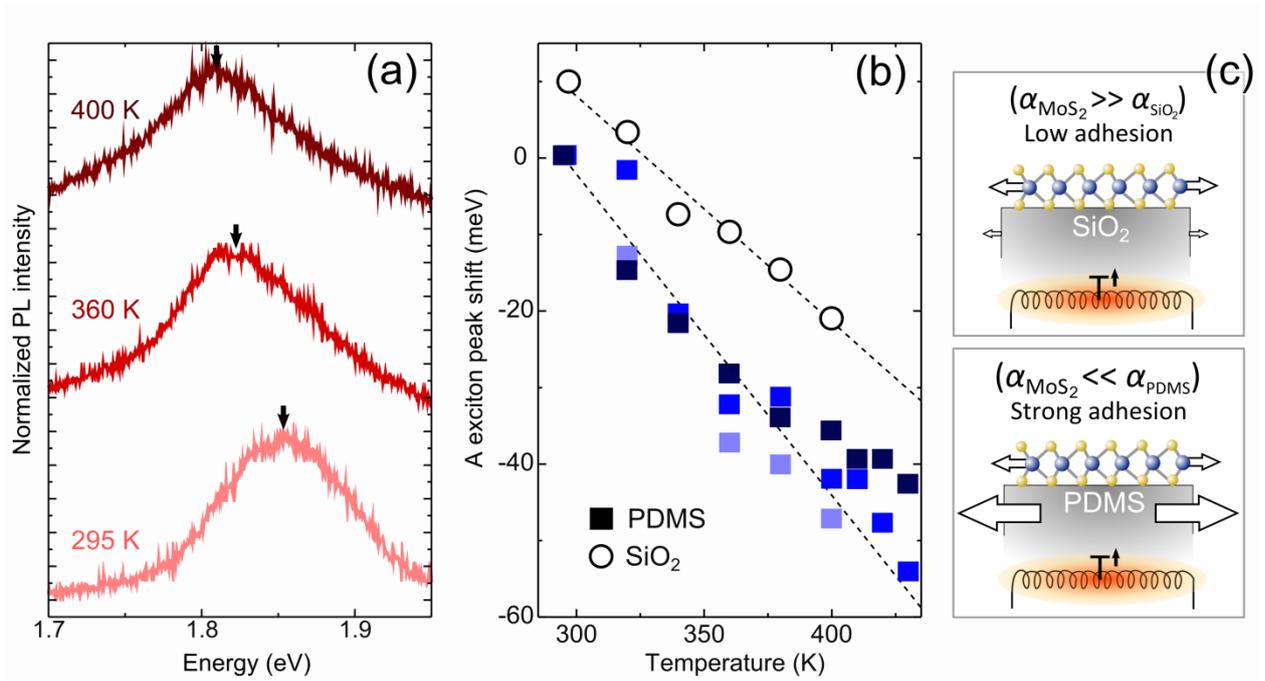

**Figure 2. Temperature-dependent photoluminescence spectra of MoS$_2$.** (a) Temperature-dependent PL spectra of MoS$_2$ on PDMS. Black arrows indicate the A exciton peak positions. (b) Temperature-dependent shift of the A exciton peak position on PDMS (squares) and SiO$_2$ (circles) substrates. Three MoS$_2$ flakes on PDMS were investigated, indicated by the different colors. The data for the SiO$_2$ was vertically shifted by 10 meV to improve visibility. The dashed lines correspond to linear fits to the data. (c) Schematic of the temperature-dependent experiment. Substrate heating causes thermal expansion of the MoS$_2$, and due to its larger thermal expansion coefficient, the PDMS substrate induces an additional biaxial strain.

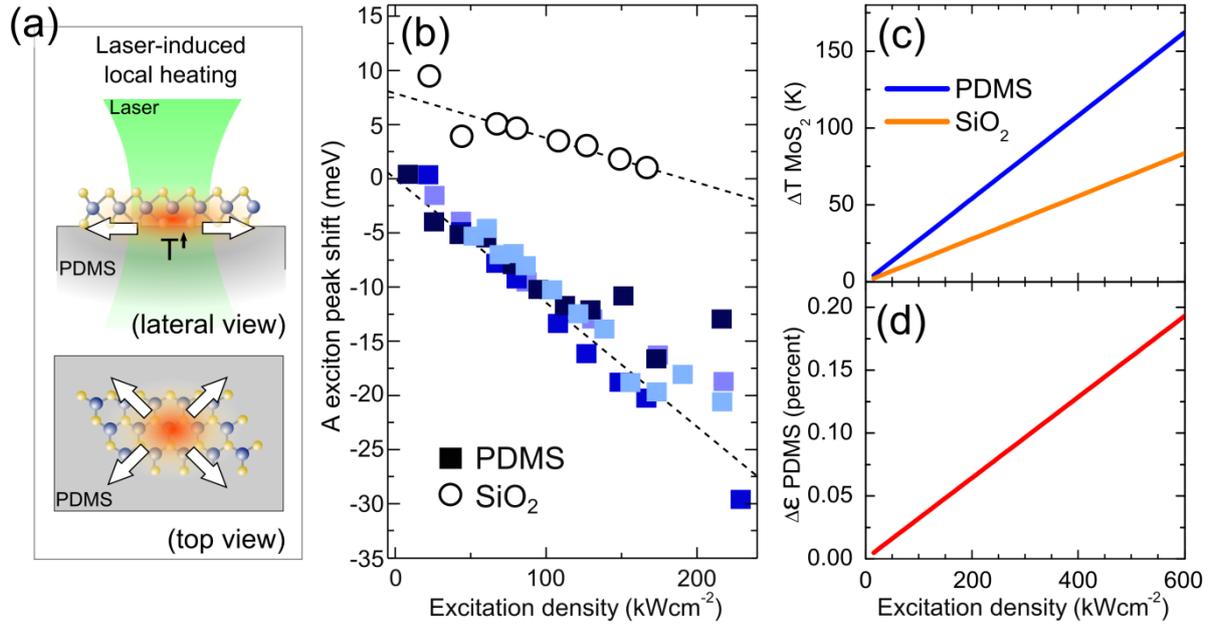

**Figure 3. Power-dependent photoluminescence spectra of MoS₂.** (a) Schematic of MoS$_2$ monolayer on top of PDMS. Focused laser excitation causes heat transfer from MoS$_2$ to PDMS and local biaxial expansion of the substrate. (b) Excitation-density-dependent shift of the A exciton peak position on PDMS (squares) and SiO$_2$ (circles) substrates. Four MoS$_2$ flakes on PDMS were investigated, indicated by the different colors. The data for the SiO$_2$ was vertically shifted by 10 meV to improve visibility. The dashed lines correspond to linear fits to the data. (c) Temperature change of MoS$_2$ flakes on PDMS and SiO$_2$ as a function of excitation density. (d) Estimated PDMS-induced strain on MoS$_2$ flake as a function of excitation density.

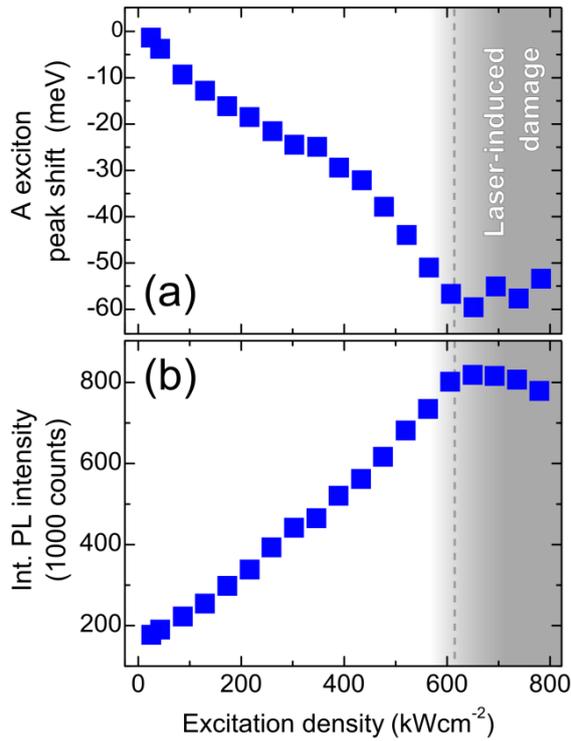

**Figure 4. Limits of laser-induced local heating.** (a) Excitation-density-dependent shift of the A exciton peak position on PDMS for large excitation densities. (b) Integrated A exciton PL intensity as a function of excitation density.

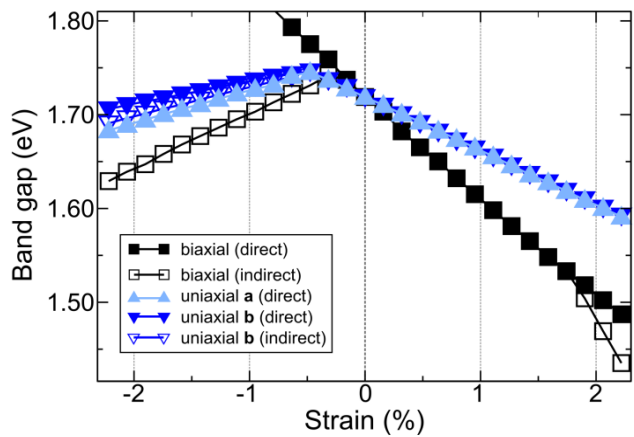

**Figure 5. Calculated strain tunability of the band gap.** The band gap evolution with respect to strain is shown for areas of compression and expansion. Biaxial strain (black symbols) is compared to uniaxial strain along the zigzag (a, light blue) and armchair (b, dark blue) directions of the lattice. Optical band gaps are denoted with full symbols. Open symbols, if present, give the fundamental (indirect) band gap.

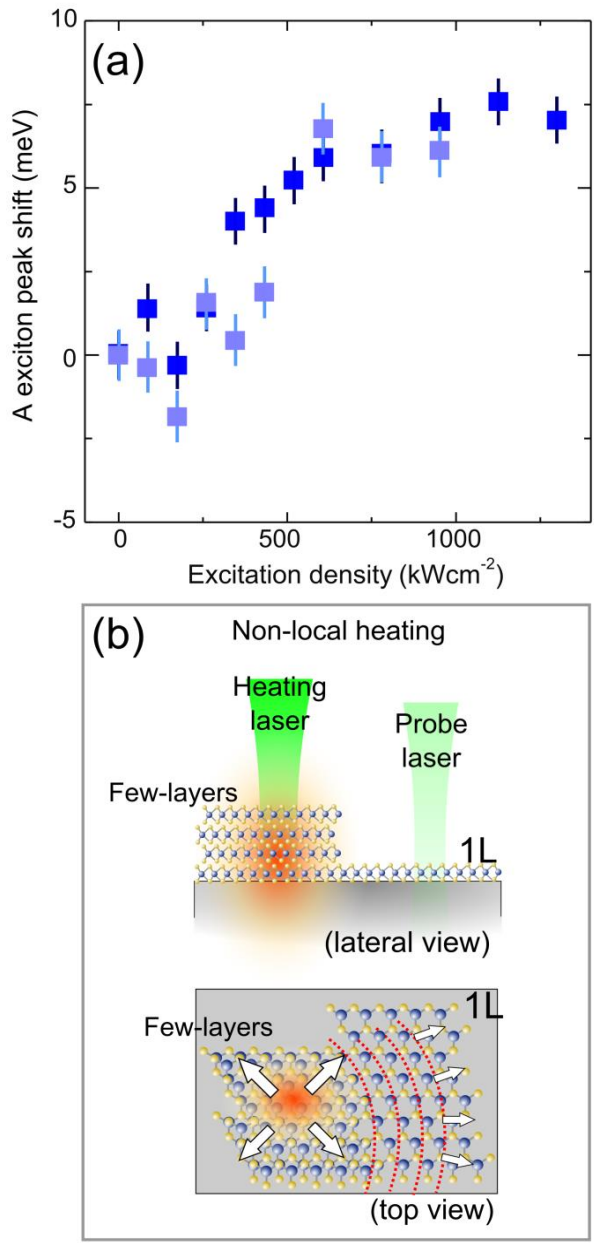

**Figure 6. Spatially separated heating effects on photoluminescence of MoS₂** (a) Shift of the A exciton peak position as a function of the excitation density of the second laser beam for the experimental geometry shown in (b), measured on two different MoS₂ flakes indicated by the different colors. (b) Schematic of experiment with spatially separated heating and strain probing. A second laser is focused onto the few-layer part of a MoS₂ flake, where it is strongly absorbed. PL is collected from the single-layer part of the same flake at a distance of 10 μm. The PDMS directly beneath the few-layer MoS₂ expands, leading to a local, near-uniaxial compression of the surrounding substrate.

# Control of biaxial strain in single-layer Molybdenite using local thermal expansion of the substrate

*Gerd Plechinger[1], Andres Castellanos-Gomez[2], Michele Buscema[2], Herre S. J. van der Zant[2], Gary A. Steele[2], Agnieszka Kuc[3], Thomas Heine[3], Christian Schüller[1], and Tobias Korn[1]\**

[1] Institut für Experimentelle und Angewandte Physik, Universität Regensburg, D-93040 Regensburg, Germany

[2] Kavli Institute of Nanoscience, Delft University of Technology, Lorentzweg 1, 2628 CJ Delft, The Netherlands.

[3] School of Engineering and Science, Jacobs University Bremen, Campus Ring 1, D-28759 Bremen, Germany.

tobias.korn@physik.uni-regensburg.de

**Supporting Information content:**

1. **Band structure calculations**

2. **Control experiment for spatially separated heating and strain probing**

3. **Material parameters**

# I. Band structure calculations

All density-functional calculations have been carried out with the help of the ADF/BAND code[1, 2] using explicit two-dimensional boundary conditions employing numerical and Slater-type basis functions of valence triple zeta quality with one polarization function (TZP)[3] with a small frozen core, the PBE-D3(BJ) functional[4, 5] which accounts for London dispersion corrections and scalar relativistic corrections. After full geometry optimizations, band structures have been calculated explicitly incorporating Spin-Orbit interactions. However, those are found to be insensitive to the strain and change by less than 10 meV for all calculations considered here. We therefore include them in the band structure plots (Figure S1), but not in the discussion regarding the band gap evolution under strain.

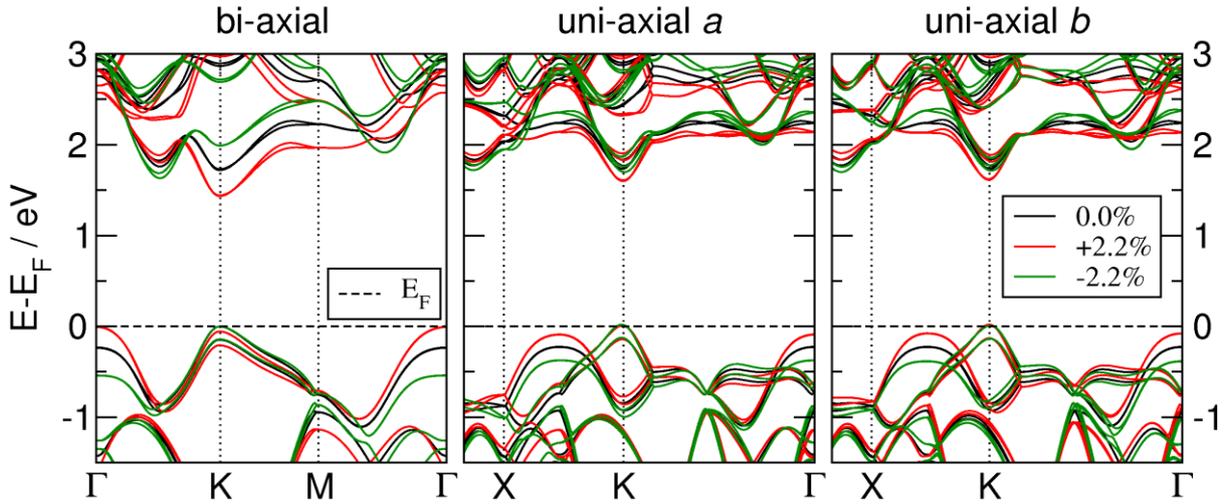

**Figure S1**. Band structure of $MoS_2$ in equilibrium geometry (black lines), strained by 2.2 % (red lines) and compressed by 2.2% (green lines), compared for bi- and uniaxial deformation of the lattice. The Fermi level is denoted as dashed line.

| Deformation | Band gap | Stretching | Compression |
|---|---|---|---|
| **bi-axial** | Direct | -105 | -120 |
| | Indirect | - | +59 |
| **uniaxial a (zigzag)** | Direct | -58 | +35 |
| **uniaxial b (armchair)** | Direct | -57 | +23 |
| | Indirect | - | +31 |

**Table S1.** Band gap evolution of $MoS_2$ under biaxial and uniaxial strain. All values are given in meV per percent of strain, with respect to increasing lattice parameters. The underlying data is visualized in Figure 5 of the manuscript.

## II. Control experiment for spatially separated heating and strain probing

To check that the uniaxial compressive strain observed in the experiment with spatially separated heating is due to localized heating of multilayer $MoS_2$, we perform a control experiment. For this, the second laser is focused onto the bare PDMS substrate, 10μm away from a monolayer $MoS_2$ flake. We utilize a weak, constant excitation density of 26 kW cm$^{-2}$ for PL excitation, and vary the excitation density of the second laser. In this measurement series, only a weak redshift of the A exciton peak occurs (about 3 meV for an excitation density of 1000 kW cm$^{-2}$), see Figure S2a. Due to the weak absorption of the laser light in the PDMS, the laser power is absorbed in a large volume of the material well below the focal plane of the second laser, as sketched in Figure S2b. Thus, focused laser illumination of the PDMS has a similar effect as global heating of the substrate, with the observed redshift corresponding to a temperature increase of about 7 K.

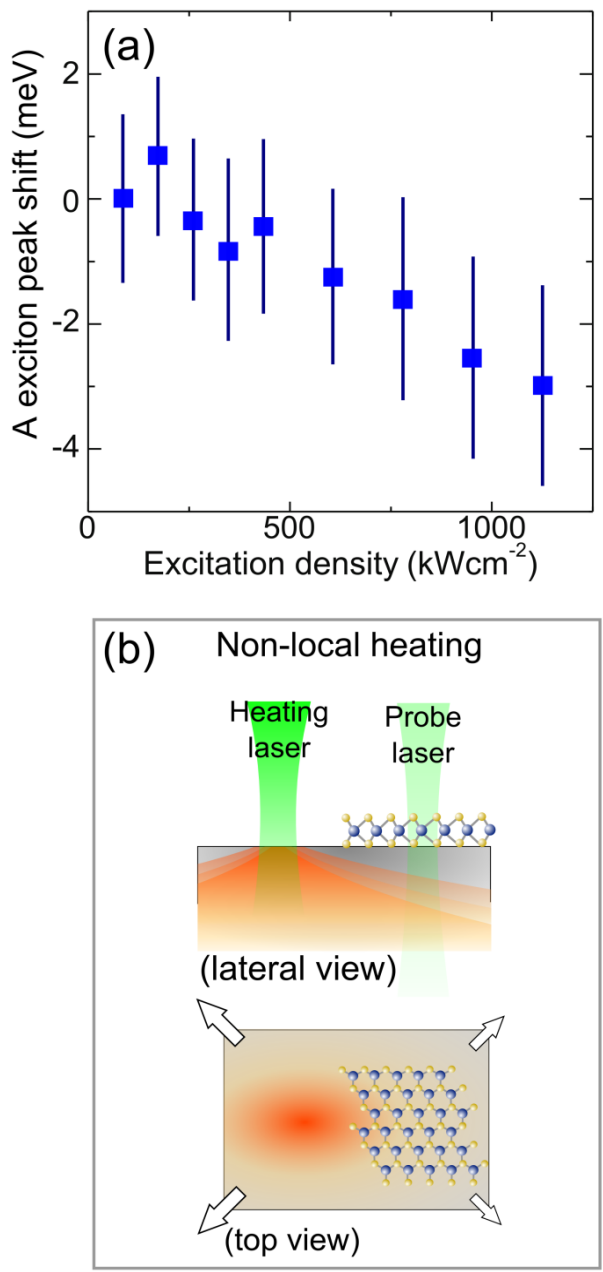

**Figure S2**. (a) Shift of the A exciton peak position as a function of the excitation density of the second laser beam for the experimental geometry shown in (b). (b) Schematic of the control experiment. A second, heating laser is focused onto the PDMS substrate 10 μm away from the monolayer $MoS_2$ flake.

### III. Material parameters

| Material | Expansion coefficient $\alpha$ ($10^{-6}$ $K^{-1}$) | Young's modulus | Thermal conductivity (W/mK) | Absorption of monolayer $MoS_2$ for green excitation | Specific heat ($Jkg^{-1}K^{-1}$) | Specific heat per volume ($Jm^{-3}K^{-1}$) |
|---|---|---|---|---|---|---|
| $SiO_2$ | $0.24^6$ | 65 $GPa^7$ | $1.3^8$ | 0.16 | 703 | 1547 |
| PDMS | $300^9$ | 360-870 $kPa^{10}$ | $0.15^{11}$ | 0.025 | 1460 | 1416 |
| $MoS_2$ | 1.9$^{12}$ (a-axis, bulk) | 330 $GPa^{13}$ | 52 (susp. Monolayer) | | | |

**Table S2.** Material parameters